\documentclass[a4paper,9pt, twocolumn]{extarticle}    %Uncomment for formatted
\usepackage[a4paper, dvips=true, scale=0.86, top=2.5cm]{geometry}  %Comment for formatting
\usepackage[T1]{fontenc}
\usepackage[note,mcite]{achemso}
\usepackage{amsmath}
\usepackage[dvips]{graphicx}
\usepackage{hyperref}
\usepackage{breakurl}

\usepackage[roman]{sublabel}

\newcommand{\ddt}[2][]{\ensuremath{\frac{\mathrm{d_{#1}}#2}{\mathrm{d}t}}}
\newcommand{\dpdt}[3][]{\ensuremath{
    \left(
      \frac{\partial_{\mathrm{#1}}#2}{\partial t}
    \right)_{#3}}}
\newcommand{\Dz}[2][r]{\ensuremath{\Delta_{\mathrm{#1}}#2^0}}
\newcommand{\scite}[1]{\textsuperscript{\mciteSubRef{#1}}}

\makeatletter
\renewcommand\section{\@startsection {section}{1}{\z@}%
       {-1.5ex \@plus -.5ex \@minus -.8ex}%
       {1.5ex \@plus.2ex \@minus .2ex}%
       {\raggedright\normalfont\large\bfseries\sffamily}}
    \renewcommand\subsection{\@startsection{subsection}{2}{\z@}%
       {-1ex\@plus -.4ex \@minus -.4ex}%
       {1ex \@plus .2ex \@minus .2ex}%
       {\raggedright\normalfont\bfseries\sffamily}}
    \renewcommand\subsubsection{\@startsection{subsubsection}{3}{\z@}%
      {2ex \@plus1ex \@minus.3ex}%
      {-1em}%
      {\normalfont\normalsize\bfseries\sffamily}}
\makeatother
\title{Energetic and Entropic Analysis of Mirror Symmetry Breaking
  Processes in a Recycled Microreversible Chemical System}

\author{\emph{Rapha\"el Plasson$^1$\thanks{\texttt{rplasson@nordita.org}},
  Hugues Bersini$^2$}\\
  $^1$Nordita, Stockholm, Sweden, $^2$IRIDIA, ULB, Brussels, Belgium.}

\begin{document}

\date{}

\onecolumn                                      %Uncomment for formatted

\thispagestyle{empty}

\maketitle

\begin{abstract}
  Understanding how biological homochirality emerged remains a
  challenge for the researchers interested in the origin of
  life. During the last decades, stable non-racemic steady states of
  nonequilibrium chemical systems have been discussed as a possible
  response to this problem. In line with this framework, a description
  of recycled systems was provided in which stable products can be
  activated back to reactive compounds. The dynamical behaviour of
  such systems relies on the presence of a source of energy, leading to
  the continuous maintaining of unidirectional reaction loops. A full
  thermodynamic study of recycled systems, composed of microreversible
  reactions only, is presented here, showing how the energy is
  transferred and distributed through the system, leading to cycle
  competitions and the stabilization of asymmetric states.
 
\end{abstract}

\twocolumn                                   %Uncomment for formatted

\section*{Introduction}

The early origin of life cannot be studied without taking account the
self-organization of chemical networks\cite{eigen-71, *kauffman-93,
  *ganti-97}, the emergence and antagonism of autocatalytic
loops\cite{farmer-86, *mossel-05}, and the onset of energy fluxes
driving the whole process\cite{morowitz-68, *morowitz-92,
  *morowitz-07, *smith-08a, *smith-08b}.  Such chemical networks are
especially interesting to understand the appearance of biological
homochirality --- still one of the most prominent problems in the
origin of life\cite{bonner-91, *avetisov-96, *podlech-01, *palyi-99,
  *palyi-02, *palyi-04, *caglioti-06, soai-95, *soai-00, *soai-03,
  *soai-06, *soai-08} --- as the destabilization of the racemic state
resulting from the competition between enantiomers and from
amplification processes concerning both competitors \cite{plasson-07}.
The onset of biological homochirality can thus be conceived as a
symmetry breaking phenomenon occurring in a nonequilibrium
autocatalytic system leading to stable non-racemic steady
states\cite{kondepudi-83, *kondepudi-90, *kondepudi-01}.

\begin{figure}[ht]
  \centering
  \includegraphics[width=8cm]{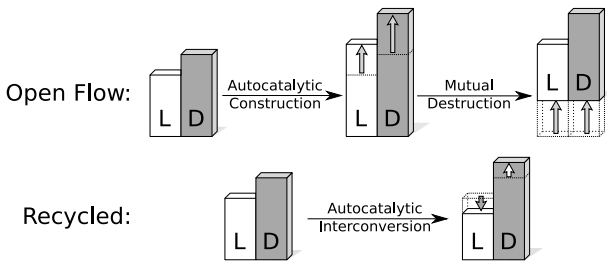}
  \caption{Schematic representation of the deracemization process in
    open-flow and recycled systems. Open-flow systems: the
    autocatalytic construction increases the absolute difference
    between L and D, and maintains the relative difference; mutual
    destruction maintains the absolute difference, and increases the
    relative difference. Recycled system: some of minor configuration
    compounds are converted to the major configuration, increasing
    both the absolute and relative differences between L and D.}
  \label{fig:frank-recyc}
\end{figure}

Recent developments in those lines are aiming at the description of
recycled systems\cite{saito-04, *viedma-05, plasson-04}, rather than
open-flow systems\cite{frank-53}. While traditional model systems are
totally open, the energy input being brought by fluxes of the chiral
subunits themselves, what we call ``recycled systems'' are instead closed to the chiral
subunits but coupled to a flux of energetic compounds. In recycled
systems, there is a constant number of chiral subunits that are driven
away from the equilibrium state by an active process allowing to
transform low-potential chiral subunits into high-potential
ones\cite{plasson-07}. This absence of addition/removal
process of chiral subunits, which are inherent to the Frank's model, implies a necessary
interconversion of chiral subunits to obtain a
change of the initial imbalance (see Fig.~\ref{fig:frank-recyc}). 
Reaction fluxes become thus necessary, going through some epimerization
reactions, directed from the minor configuration towards the major one,
that is, in a direction opposed to the thermodynamic one.

While previous studies have focused on the kinetics
aspects\cite{plasson-04}, the purpose of this work is to understand
how such chemical dynamic systems can be conceived still obeying the
thermodynamic laws. Our conceptual framework will lay an important
emphasis on entropic exchanges\cite{prigogine-62a, *groot-69,
  *kondepudi-98, yoshida-90, *irvin-88, *dutt-90, *kondepudi-08}. This approach
requires to take into account the microreversibility of all reactions,
even if they turn out to be quasi-irreversible, and to explicit the source of
energy. These points can be neglected in a kinetic
study\cite{plasson-08a, *blackmond-08b}, but are necessary to compute energetic
quantities.

Departing from there, a nonequilibrium Onsager's triangle of
reactions\cite{onsager-31} will be analyzed to point out how chemical
energy can be transferred into a chemical system and maintain a
unidirectional reaction cycle. This will lead to the description of
nonequilibrium systems in terms of the reaction fluxes (that is, the
\emph{dynamics} of the systems), rather than the concentrations (that
is, their \emph{static} parameters).  Finally, the APED system
(Activation-Polymerization-Epimerization-Depolymerization
system\cite{plasson-04}) will be analyzed, showing how a recycled
system of chiral subunits can be maintained in a non-racemic state by
consuming chemical energy, and by driving this energy towards reaction
cycles of autocatalytic inversion of chiral centers. An energetic and
entropic analysis will show how this source of energy can be
efficiently consumed.

\section*{General Relations}

%\paragraph{Description:}

Let us consider a system constituted by $r$ chemical reactions $R_j$,
involving $n$ compounds $X_i$. For each compound $X_i$, $n$ fluxes of exchange of
matter $\varPhi_i$ can be established with the surrounding:
\sublabon{equation}
\begin{eqnarray}
    %\xrightarrow{f_i}
    \varPhi_i: && \stackrel{\varphi_i^\mathrm{e}}{\rightarrow}
     X_i\\
    R_j:& \displaystyle\sum_{i=1}^n \nu^-_{i,j} X_i 
    %\xrightleftharpoons[\quad k_{-j}\quad]{k_j} &
    & \underset{k_{-j}}{\stackrel{k_j}{\rightleftharpoons}} 
    \displaystyle\sum_{i=1}^n \nu^+_{i,j} X_i  .
\end{eqnarray}
\sublaboff{equation}
The exchange flux\scite{iupac-chemflux} $\varphi_i^\mathrm{e}$ is
positive for incoming compounds, negative for outgoing compounds, and
zero for non-exchanged compounds\cite{notations,*naught-97,
  *iupac-chemflux, *iupac-rateapp, *iupac-microrev,
  *iupac-detailedbalance}. The net stoichiometric coefficients are:
\begin{eqnarray}
  \nu_{i,j}&=&\nu^+_{i,j}-\nu^-_{i,j} .
\end{eqnarray}
If $\nu^+_{i,j}=0$ and $\nu^-_{i,j}\neq 0$, $X_i$ is a reactant of the
reaction $R_j$. If $\nu^+_{i,j} \neq 0$ and $\nu^-_{i,j}= 0$, $X_i$ is a
product. If $\nu^-_{i,j}=\nu^+_{i,j}\neq0$, $X_i$ is a catalyst. If
$\nu^+_{i,j}\neq0$, $\nu^-_{i,j}\neq 0$ and $\nu_{i,j} \neq 0$, $X_i$
is an autocatalyst. If $\nu^-_{i,j}=\nu^+_{i,j}=0$, $X_i$ does not
participate in the reaction $R_j$.

The rate of appearance\scite{iupac-rateapp} $v_i$ in each compound
$X_i$ is the resultant from exchange fluxes and reaction fluxes (See
Appendix from Eq.~\ref{eq:def-flux-start} to Eq.~\ref{eq:def-flux-end}
for more details):
\begin{eqnarray}
  v_i  &=& \varphi_i^\mathrm{e} + \sum_{j=1}^r \nu_{i,j}\varphi_j   .
\end{eqnarray}

The purpose of this study is to describe the circulation of matter and
energy fluxes inside the system. Like appropriately underlined by
Blackmond and Matar\cite{blackmond-08, plasson-08a}, all reactions
must be described as microreversible\scite{iupac-microrev}. This means
that in the equilibrium state, each reactions must obey detailed
balance\scite{iupac-detailedbalance}, implying that all independent
fluxes $\varphi_j$ are zero (see the Appendix part ``Microreversibility
and detailed balancing'' for more details). As a consequence, direct
and indirect kinetic constants ($k_j$ and $k_{-j}$) must be
consistently related with the thermodynamic equilibrium constant $K_j$.

An elegant way to automatically take into account all these
relationships in complex chemical networks was described by Peusner
and Mikulecky\cite{peusner-85, *mikulecky-01}. An analogy between
chemical networks and electric networks can be done, in which case a
chemical reaction can be described by the resistive onset of chemical
fluxes between reactant and product ``nodes''. This can be done by
defining one characteristic parameter per element: the $\Gamma_j$ for
each reaction, and the standard constant of formation
$K_{\mathrm{f},i}=e^{-\frac{\Dz[f]{G}_i}{RT}}$ for each compound. In
that context, a given state of the system is totally specified by one
variable for each element: the flux $\varphi_j$ of each reaction, and
the potential $V_i$ of each compound.

The parameter $V_i$, characteristic of the state of the compound $X_i$
is defined by:
%\sublabon{equation}
\begin{eqnarray}
  V_i&=&\frac{x_i}{K_{\mathrm{f},i}}\\
  &=&e^{\frac{\mu_i}{RT}}  .
\end{eqnarray}
%\sublaboff{equation}
$V_i$ is the activity of compound $X_i$ (taken equal to its
concentration $x_i$) corrected by its relative stability (reflected by
its standard constant of formation $K_{\mathrm{f},i}$).

The parameter $\Gamma_j$, characteristic of the reaction $R_j$ is
defined by (See Appendix from Eq.~\ref{eq:def-gamma-start} to
Eq.~\ref{eq:def-gamma-end} for more details):
\begin{eqnarray}
  \Gamma_j  &=&k_{-j}\displaystyle\prod_{i=1}^n K_{\mathrm{f},i}^{\nu^+_{i,j}}=
  k_j\displaystyle\prod_{i=1}^n K_{\mathrm{f},i}^{\nu^-_{i,j}}.\label{eq:detailed-bal}
\end{eqnarray}
$\Gamma_j$ reflects the general speed of the reaction $R_j$ in both
directions. The effective constant rate is then equal to $\Gamma_j$
(high values reflecting fast reaction) divided by the standard
constants of formation of the reactants (stable reactants leading to
slow reactions, unstable reactants to fast reactions).

The expression of the chemical flux for each reaction $R_j$ then
becomes (Appendix Eq.~\ref{eq:demo-ohm}):
%\sublabon{equation}
\begin{eqnarray}
  \varphi_j^-&=&\Gamma_j \prod_{i=1}^n V_i^{\nu^-_{i,j}}  , \\
  \varphi_j^+&=&\Gamma_j \prod_{i=1}^n V_i^{\nu^+_{i,j}}  , \\
  \varphi_j&=&\Gamma_j
  \left(
    \prod_{i=1}^n V_i^{\nu^-_{i,j}}
    -
    \prod_{i=1}^n V_i^{\nu^+_{i,j}}
  \right)   \label{eq:ohm-reac} .
\end{eqnarray}
%\sublaboff{equation}
This equation is equivalent to a chemical ``Ohm's law'', describing
the onset of a chemical flux generated by differences of chemical
potential between reactants and products.

The entropy production by one reaction (Appendix
Eq.~\ref{eq:entrop-prod-ann}) is:
%\sublabon{equation}
\begin{eqnarray}
  \sigma_j&=&-R\varphi_j\ln
  \left(
    \prod_{i=1}^n V_i^{\nu_{i,j}}
  \right)\label{eq:entrop-prod}   ,\\
  &=&R(\varphi_j^+-\varphi_j^-)\ln\frac{\varphi_j^+}{\varphi_j^-} .
\end{eqnarray} 
%\sublaboff{equation}
This is equivalent to an energy dissipation, $\sigma_j$ being positive
whatever the actual direction of the flux. The total entropy
production of the network is:
\begin{eqnarray}
  \sigma^\mathrm{i}&=& \sum_{j=1}^r \sigma_j       .
\end{eqnarray}
The entropy of exchange (Appendix Eq.~\ref{eq:entrop-exch-ann}) is:
\begin{eqnarray}
  \sigma^\mathrm{e}  &=&R\sum_{i=1}^n  \varphi_i^\mathrm{e}\ln
  V_i   \label{eq:entrop-exch}. 
\end{eqnarray} 
A positive value of $\sigma^\mathrm{e}$ indicates an increase of
internal energy $T\sigma^\mathrm{e}$ due to the matter exchanges (and
symmetrically, a negative value indicates a decrease).

The entropy balance can be computed by:
\begin{eqnarray}
  \label{eq:ent-bal}
  \sigma_{bal}&=&\sigma^\mathrm{e} - \sigma^\mathrm{i} .
\end{eqnarray}
The global energy exchange is $\varepsilon=T\sigma_{bal}$. When
$\sigma_{bal}>0$, the system globally consumes energy. When
$\sigma_{bal}<0$, the system globally releases energy.

A steady state corresponds to:
\begin{eqnarray}
  \varphi_i^\mathrm{e}&=&- \sum_{j=1}^r \nu_{i,j} \varphi_j   .
\end{eqnarray}
The exchanges with the surrounding can compensate an unbalance of the
reactions, maintaining continuous fluxes $\varphi_j$. In this case,
$\sigma_{bal}$ is zero, that is $\sigma^\mathrm{e}=
\sigma^\mathrm{i}$. The system exactly dissipates what it receives,
maintaining constant its internal state. The maintaining of this state
is active, and is different from the unique equilibrium state. The
matter fluxes $\varphi^\mathrm{i}$ imply a continuous creation of
entropy during the transformation. The matter flux maintains the
reaction, allowing it to be continuously performed. 

The equilibrium state is a special case of steady state where each
$\varphi_j=0$ and $\varphi_i^\mathrm{e}=0$, in which case the absence
of exchanges leads to detailed balance\scite{iupac-detailedbalance}.
There are no exchange of any sort with the surrounding, and no entropy
is produced: $\sigma^\mathrm{e}$, $\sigma_{bal}$ and all $\sigma_j$
are zero.

\subsection*{Nonequilibrium Onsager's Triangle}

Coupling a chemical system to a source of energy leads to the
spontaneous formation of fluxes, possibly leading to positive feedback
and self-organization\cite{morowitz-68, *morowitz-92, *morowitz-07,
  *smith-08a, *smith-08b}. Such couplings are actually ubiquitous in
all biological systems, where many endergonic biochemical reactions
are coupled to the hydrolysis of ATP. Some nonequilibrium abiotic
systems also function on the basis of chemical energy
transfer\cite{field-72, *nicolis-73}. These nonequilibrium systems are
characterized by a closed system of given chemical compounds,
performing unidirectional loops through the consumption of an excess
(or clamped) of ``fuel'' molecule. The canonical following example
describes how energy can generate and maintain an unidirectional cycle
of reaction.

\subsubsection*{Description of the System.}

\begin{figure}[ht]
  \centering
  \includegraphics[width=8cm]{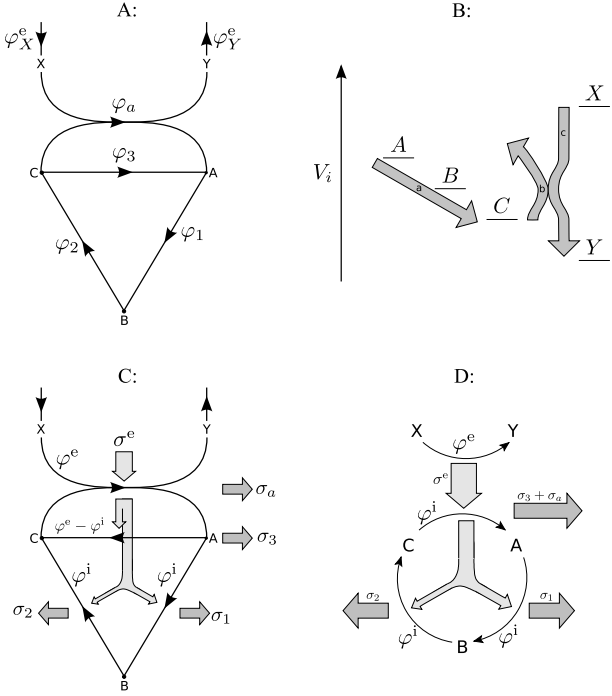}
  \caption{A: Schematic representation of the reaction network
    relative to a nonequilibrium Onsager triangle. B: Steady state
    potential of the
    system\cite{plasson-08a}\textsuperscript{.a}. \texttt{a},
    spontaneous transformation from $A$ to $C$; \texttt{b}, activated
    transformation from $C$ to $A$; \texttt{c}, spontaneous
    transformation from $X$ to $Y$, coupled to the activated
    transformation \texttt{b}, the process \texttt{c} forcing the
    process \texttt{b} to compensate for the process \texttt{a}. C:
    Detailed distribution of the entropic fluxes inside the whole
    network. D: Simplified description of the system, emphasizing the
    coupling between the external compounds $X$,$Y$ and the internal
    compounds $A$,$B$,$C$.}
  \label{fig:triangle}
\end{figure}

At equilibrium, a reaction loop can't be subject to a unidirectional
flux of reactions, as all reactions become balanced in
detail\cite{onsager-31, boyd-77}. Nonequilibrium steady states rely on
energy consumption dissipated by entropy production, leading to cyclic
processes\cite{klein-55}. In order to understand what happens in these
steady state systems, let us consider the following system:
\begin{equation}
  \begin{array}{rclcc}
    &
    %\xrightarrow{\varphi_{e}}
    \stackrel{\varphi_X^\mathrm{e}}{\rightarrow}
    & X&\qquad;\qquad&\varphi^{\mathrm{e}}_X\\
    C + X &
    %\xrightleftharpoons[k_{-a}]{k_a}
    \underset{k_{-a}}{\stackrel{k_a}{\rightleftharpoons}}
    & A + Y&\qquad;\qquad&\varphi_a\\
    Y & 
    %\xrightarrow{\varphi_{e}}
    \stackrel{\varphi_Y^\mathrm{e}}{\rightarrow}
    &&\qquad;\qquad&\varphi_Y^\mathrm{e}\\
    A & 
    %\xrightleftharpoons[k_{-1}]{k_1}
    \underset{k_{-1}}{\stackrel{k_1}{\rightleftharpoons}}
    & B&\qquad;\qquad&\varphi_1\\
    B & 
    %\xrightleftharpoons[k_{-2}]{k_2} 
    \underset{k_{-2}}{\stackrel{k_2}{\rightleftharpoons}}
    & C&\qquad;\qquad&\varphi_2\\
    C & 
    \underset{k_{-3}}{\stackrel{k_3}{\rightleftharpoons}}
    %\xrightleftharpoons[k_{-3}]{k_3} 
    & A&\qquad;\qquad&\varphi_3    
  \end{array}\label{eq:def-onsag}   .
\end{equation}
The whole network is represented in fig.~\ref{fig:triangle}A.
Fluxes of $X$ and $Y$ force the reaction from $C$ to $A$, and thus
maintain the system $\{A,B,C\}$ out of equilibrium. 
The reaction fluxes can be expressed as:
\sublabon{equation}
\begin{eqnarray}
  \varphi_a&=& \Gamma_a
  \left(
    V_cV_x - V_aV_y
  \right)\label{eq:ohm0}\\
  \varphi_1&=& \Gamma_1 \left(
    V_a - V_b
    \right)\label{eq:ohm1}\\
  \varphi_2 &=& \Gamma_2
  \left(
    V_b - V_c
  \right) \label{eq:ohm2}\\
  \varphi_3 &=& \Gamma_3
  \left(
    V_c - V_a 
  \right)\label{eq:ohm3} .
\end{eqnarray}
\sublaboff{equation}

%\sublaboff{equation}%

\subsubsection*{Steady State.}
In the steady state, we have: 
%\sublabon{equation}
\begin{eqnarray}
v_b=0=\varphi_1-\varphi_2 & \Rightarrow &
\varphi_1=\varphi_2=\varphi^\mathrm{i}   , \\
v_x=0=\varphi_X^\mathrm{e}-\varphi_a & \Rightarrow &
\varphi_X^\mathrm{e}=\varphi_a=\varphi^\mathrm{e} , \\
v_Y=0=\varphi_Y^\mathrm{e}+\varphi_a & \Rightarrow &
\varphi_Y^\mathrm{e}=-\varphi_a=-\varphi^\mathrm{e} , \\
v_a=0=\varphi_a-\varphi_1+\varphi_3 & \Rightarrow &
\varphi_3=\varphi^\mathrm{i}-\varphi^\mathrm{e} \label{eq:node} .
\end{eqnarray}
%\sublaboff{equation}
Note that this comes down to apply the equivalent of the Kirchhoff's
current law for a steady state chemical system\cite{schnakenberg-76}.

The relationship between the several fluxes is thus:
%\sublabon{equation}
\begin{alignat}{2}
  \left(
    \Gamma_1^{-1} + \Gamma_2^{-1}
  \right) \varphi^\mathrm{i} &= V_a - V_c &
  \qquad&((\ref{eq:ohm1})+(\ref{eq:ohm2}))
  \label{eq:2p3}   ,\\
   \Gamma_3^{-1} \varphi_3 &= -  \left(
    \Gamma_1^{-1} + \Gamma_2^{-1}
  \right) \varphi^\mathrm{i} &
  \qquad& ((\ref{eq:ohm3}) = -(\ref{eq:2p3})) 
  \label{eq:i0i}   ,\\ 
  \varphi^\mathrm{i}
  &=\frac{1}{1+\frac{\Gamma_3}{\Gamma_1}+\frac{\Gamma_3}{\Gamma_2}}\varphi^\mathrm{e}&
  \qquad& ((\ref{eq:node}) \&
  (\ref{eq:i0i}))  \label{eq:repart-i}   .
\end{alignat}
%\sublaboff{equation}
For a positive exchange flux $\varphi^\mathrm{e}$,
$\varphi^\mathrm{i}$ is positive and $\varphi_3$ is negative. There is
a net continuous transformation around the cycle, initiated from $C$
to $A$, transmitted from $A$ to $B$ and from $B$ to
$C$. $\varphi^\mathrm{i}=\varphi^\mathrm{e}+\varphi_3$ is the
effective flux induced in the cycle. $\varphi_3$ is a ``leak'' of the
activation flux. The induced cycle flux is equal to the exchange flux
($\varphi^\mathrm{i} \simeq \varphi^{e}$) when $\Gamma_3\ll \Gamma_1$
and $\Gamma_3\ll \Gamma_2$, that is, when the kinetics of reaction 3
can be neglected compared to the kinetics of reactions 1 and 2. The
fraction of flux that is induced in the cycle is independent of the
kinetics of the activation reaction, which only plays a role for the
intensity of the steady state exchange flux (efficient activation
reactions imply high exchange fluxes).

At equilibrium, all the $V_i$ are equal while in the nonequilibrium
steady state, the exchange of chemical energy allows to maintain
$V_a>V_b$ (Eq.~\ref{eq:ohm1}),  $V_b>V_c$ (Eq.~\ref{eq:ohm2}),
$V_a>V_c$ (Eq.~\ref{eq:ohm3}), and $V_cV_x>V_aV_y$ (Eq.~\ref{eq:ohm0})
which comes down to:
\begin{equation}
  \frac{V_x}{V_y}>\frac{V_a}{V_c}>1 \label{eq:cond-cycle}      ,
\end{equation}
and thus $V_x>V_y$. Note here that although $A$ is of higher potential than $C$, there is a net conversion
from $C$ to $A$ as $\varphi_a+\varphi_3= \varphi^\mathrm{i}>0$. The transfer of chemical energy
allows the recycling of low potential compound $C$ back to high
potential compound $A$, counteracting the spontaneous evolution in the
opposite direction (see fig.~\ref{fig:triangle}B).

\subsubsection*{Entropic Analysis.}

The different entropy contributions produced or exchanged during the
processes are:
\sublabon{equation}
\begin{eqnarray}
  \sigma^\mathrm{e} &=&R\varphi^\mathrm{e}\ln\frac{V_x}{V_y}\\
  \sigma_{a}        &=&R\varphi^\mathrm{e}\ln\frac{V_cV_x}{V_aV_y}\\
  \sigma_{1}        &=&R\varphi^\mathrm{i}\ln\frac{V_a}{V_b}\\
  \sigma_{2}        &=&R\varphi^\mathrm{i}\ln\frac{V_b}{V_c}\\
  \sigma_{3}
  &=&R(\varphi^\mathrm{e}-\varphi^\mathrm{i})\ln\frac{V_a}{V_c} .
\end{eqnarray}
\sublaboff{equation} 
The incoming energy, due to matter exchange, is shared by the four
reactions. These exchanges are represented in
fig.~\ref{fig:triangle}C.

\subsubsection*{Simplified Description.}

The full chemical system can be divided in two subsystems: the
internal system, only subject to internal reaction loops, and the
exchange system, involving compounds that can be exchanged with the
surrounding, and thus subject to linear chains of reactions connected
to external sources of matter. These two subsystems can be easily
determined from the laws of mass conservation. The conserved moieties
can be computed from the left nullspace of the stoichiometric matrix
$\boldsymbol{\nu}=[\nu_{i,j}]$ of the reaction
network\cite{schuster-95} (the zeros of the stoichiometric matrix are
replaced by dots for better readability): {\footnotesize%
\begin{eqnarray}
  %\footnotesize
  \boldsymbol{\nu} =\begin{bmatrix}
    -1          & \cdot           & +1     & +1\\
    +1          & -1              &  \cdot &  \cdot\\
     \cdot      & +1              & -1     & -1\\
     \cdot      &  \cdot          &  \cdot & -1\\
     \cdot      &  \cdot          &  \cdot & +1
  \end{bmatrix} & \Longrightarrow &
  %\footnotesize
  \mathrm{Null}(\boldsymbol{\nu})=
  \bordermatrix{%
     & &\cr
    A&1&\cdot\cr
    B&1&\cdot\cr
    C&1&\cdot\cr
    X&\cdot&1\cr
    Y&\cdot&1
  }   .
\end{eqnarray}}
The first vector of the left null space base gives a pool of mass
conservation for $\{A,B,C\}$, and the second vector for $\{X,Y\}$.
$X$ and $Y$ compounds are involved in matter exchanges, so that their
concentrations are ruled by the external conditions. No flux of $A$,
$B$ or $C$ is possible, so the total concentration $a+b+c$ is
constant (i.e. the system $\{A,B,C\}$ can be considered as closed, in
communication with the open system $\{X,Y\}$).

The subsystem $\{A,B,C\}$ is composed of a single unidirectional loop
of transformations from $A$ to $B$, then to $C$, and back to $A$,
performed at a constant rate $\varphi^\mathrm{i}$ (see
Fig.~\ref{fig:triangle}D). The subsystem $\{A,B,C\}$ acts as a closed
system, maintained in a nonequilibrium activity through its
coupling with the open-flow system $\{X,Y\}$. The incoming chemical
energy $T\sigma_e$ is transferred from $\{X,Y\}$ to $\{A,B,C\}$, and
dissipated by continuous entropy creation:
%\sublabon{equation}
\begin{eqnarray}
  \sigma^\mathrm{i} &=&\sigma_a+\sigma_1+\sigma_2+\sigma_3\\
  &=&\sigma^\mathrm{e}  .\label{eq:dissip-triangle}
\end{eqnarray}
%\sublaboff{equation}

Provided the system $\{X,Y\}$ is such that the concentrations $x$ and $y$
can be considered as constant (e.g. in the case of a large reservoir
of energy compared to a small system $\{A,B,C\}$), the system
$\{A,B,C\}$ is mathematically strictly equivalent to a closed system,
in which we have first order reactions in $A$, $B$ and $C$. The
corresponding \emph{apparent} kinetic rates would seem not to respect
the microreversibility, as they implicitly depend on the
concentrations $x$ and $y$; each time such theoretical chemical
network is built\cite{cera-89,*basu-99,*mauksch-07}, it is fundamental
to realize that there is a hidden source of energy in the system.

\subsubsection*{Time Evolution.}

Numerical integration of the several subsets of the system were
performed using Xppaut\cite{xppaut}, with
$K_{\mathrm{f},A}=1~\mathrm{M}$, $K_{\mathrm{f},B}=2~\mathrm{M}$,
$K_{\mathrm{f},C}=4~\mathrm{M}$, $K_{\mathrm{f},X}=1~\mathrm{M}$,
$K_{\mathrm{f},Y}=10~\mathrm{M}$, $\Gamma_1=1~\mbox{M.s}^{-1}$,
$\Gamma_2=2~\mbox{M.s}^{-1}$, $\Gamma_3=4~\mbox{M.s}^{-1}$ and
$\Gamma_a=40~\mbox{M.s}^{-1}$. This corresponds to
$k_1=k_2=k_3=1~\mbox{s}^{-1}$, $k_a=10~\mbox{M}^{-1}.\mbox{s}^{-1}$,
$k_{-1}=k_{-2}=0.5~\mbox{s}^{-1}$, $k_{-3}=4~\mbox{s}^{-1}$,
$k_{-a}=4~\mbox{M}^{-1}.\mbox{s}^{-1}$, $K_1=\frac{b_{eq}}{a_{eq}}=2$,
$K_2=\frac{c_{eq}}{b_{eq}}=2$ and $K_a=\frac{y_{eq}}{x_{eq}}=10$.
Initial conditions were $a_0=3$~M, $b_0=2$~M, $c_0=2$~M. $x$ and
$y=10^{-5}$~M were maintained constant for each case, corresponding to
the presence of a large reservoir of $X$ and $Y$ connected to the
internal system. Then, $x$ was varied as a parameter, allowing to tune
the quantity of incoming energy.

\begin{figure}[b]                               % Uncomment for formatted
%\begin{figure}[hbt]                              % Comment for formatted
  \centering
  \includegraphics[width=8cm]{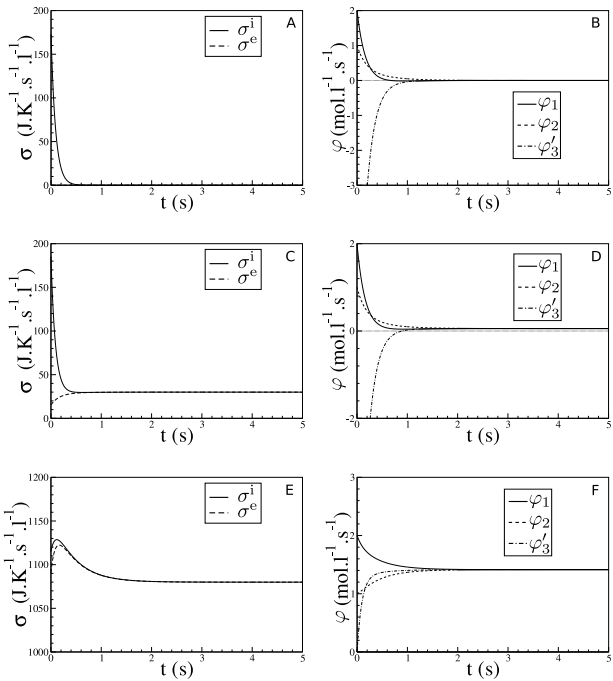}            %Uncomment for formatted 
  \caption{Time evolution of entropy productions (A, C, E) and
    chemical fluxes (B, D, F) in nonequilibrium Onsager's triangles
    as defined in Eq.~\ref{eq:def-onsag}. A and B: $x=0$~M, C and D:
    $x=0.01$~M, E and F: $x=0.5$~M. See text for the other numerical
    values.}
\label{fig:time-loop}
\end{figure}

The evolution of the different fluxes of transformation $\varphi_1$,
$\varphi_2$ and $\varphi_3'=\varphi_3+\varphi_a$ (i.e. the total
fluxes of transformation from $A$ to $B$, from $B$ to $C$, and from
$C$ to $A$) was examined. Under these conditions, three different kind
of evolution were observed (see Fig.~\ref{fig:time-loop}):
\begin{description}
\item[Equilibrium] ($x=0$): There is no energy source.  The system
  initially dissipates its excess of internal energy, to evolve
  spontaneously towards the equilibrium state ($a=1$~M, $b=2$~M,
  $c=4$~M) where the entropy production is zero (i.e. the system
  reaches a maximum of entropy). As only spontaneous reactions are
  possible in absence of coupling, the reaction runs from $A$ to $B$
  and from $B$ to $C$ ($\varphi_1>0$, $\varphi_2>0$) and from $A$ to
  $C$ ($\varphi'_3<0$). When the equilibrium is reached, all the
  fluxes are zero: all reactions obey detailed balance.
\item[Close to equilibrium] ($x=0.01$): There is a weak source of
  energy. The evolution is very similar, the system evolves towards a
  steady state that is very close to the equilibrium state ($a=1.07$~M,
  $b=2.01$~M, $c=3.92$~M). The major difference is observed after some
   time. The entropy production does not go to zero, but diminishes
  until reaching the exchange entropy, so that the incoming energy is
  totally dissipated. The system thus reaches a minimum of entropy
  production. The fluxes also follow a similar evolution as previously
  ($\varphi_1>0$, $\varphi_2>0$, $\varphi'_3<0$), except that  they now reach
  a non zero positive flux $\varphi^\mathrm{i}$, $\varphi'_3$ reversing its direction when
  approaching the steady state $\varphi_1=\varphi_2=\varphi_3'>0$.
\item[Far from equilibrium] ($x=0.5$): There is a strong source of
  energy. This time, the entropy production increases before decreasing
  towards the steady state  ($a=2.6$~M,
  $b=2.4$~M, $c=2.0$~M). During the whole time, the
  entropy production remains actually quite close to the exchange
  entropy. A unidirectional cycle of reaction is almost
  instantaneously obtained in the system, ($\varphi_1>0$, $\varphi_2>0$,
  $\varphi'_3>0$). The far-from-equilibrium system is ruled almost only by
  the energy flux.
\end{description}

\begin{table*}[btp]
  \centering
  \caption{Entropy productions, chemical fluxes and compounds
    concentration at the steady state in nonequilibrium 
    Onsager triangle for different incoming
    fluxes.\label{tab:triangle-entprod}}
  %\small{
    \begin{tabular}{ccccccccccc}
      $x$    & $\sigma^\mathrm{e}$               & $\sigma_1$           & $\sigma_2$           & $\sigma_3$             & $\sigma_a$                      & $\varphi^\mathrm{i}$ & $\varphi^\mathrm{e}$ & $a$             & $b$                             & $c$\\
      (M)    &  (J.K$^{-1}$.s$^{-1}$.l$^{-1}$)    & (\%)                 &  (\%)                &       (\%)             &  (\%)                           &  (M.s$^{-1}$)        &  (M.s$^{-1}$)        &       (M)       &  (M)                            &  (M)\\
      \hline
      $0$    & $0$                              & ---                  & ---                  & ---                    & ---                             & $0$                  & $0$                  & $1$             & $2$                             & $4$\\
      $0.01$ & $30$                             & $0.08$               & $0.04$               & $0.8$                  & $99.1$                          & $0.056$              & $0.392$              & $1.07$          & $2.01$                          & $3.92$\\
      $0.1$  & $318$                            & $0.5$                & $0.3$                & $4.6$                  & $94.6$                          & $0.475$              & $3.32$               & $1.6$           & $2.1$                           & $3.3$\\
      $0.5$  & $1080$                           & $0.8$                & $1.0$                & $10.8$                 & $87.3$                          & $1.41$               & $9.90$               & $2.6$           & $2.4$                           & $2.0$\\
      $1$    & $1511$                           & $0.9$                & $1.40$               & $14.0$                 & $83.6$                          & $1.88$               & $13.15$              & $3.1$           & $2.5$                           & $1.3$\\
      \hline
    \end{tabular}
  %}
\end{table*}

The distribution of the fluxes for different values of $x$ is given in
Tab.~\ref{tab:triangle-entprod}. The exchange entropy grows with the
availability of the energy source $X$. It can be seen that the
majority of the energy is dissipated by the activation reaction, only
a small fraction of energy being transmitted to the cycle (from 1\% to
16\% in the performed experiments). In accordance with
Eq.~\ref{eq:repart-i}, it can be checked that the forced circular flux
$i$ always represents 14\% of the $XY$ exchange flux
$\varphi^\mathrm{e}$, whatever the value of the incoming flux.

\section*{The APED Model}

\subsection*{Energy Diagram}

The APED model describes a system based on the activation of monomers
(that can typically be amino acids), the polymerization of activated
monomers with unactivated monomers or polymers, the epimerization of
one end-residue of the polymers, and the depolymerization of the
polymers\cite{plasson-07, plasson-04}. All these reactions take place
at the same time, in a system that is closed in terms of monomer
residues (they never enter or leave the system). The activation
reaction is coupled to the consumption of chemical energy, maintaining
the system in a nonequilibrium state. It was shown that such a system
can lead to a stable non-racemic steady state, with the approximation
that most of the reactions are irreversible\cite{plasson-04,
  plasson-08a}.

\begin{figure}[ht]
  \centering
  \includegraphics[width=8cm]{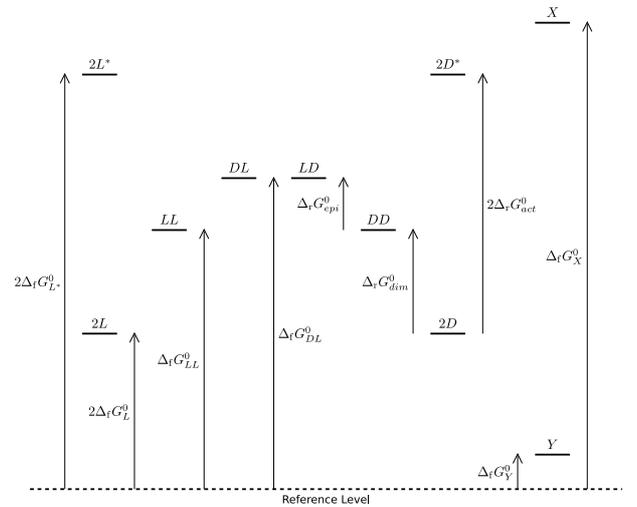}
  \caption{Thermodynamic diagram of the different compounds engaged in
    the APED system.}
\label{fig:energ-level}
\end{figure}

Let us analyze the behavior of the APED system, limited to dimers for
the sake of simplicity, appropriately taking into account 
the microreversibility of all
the reactions \cite{blackmond-08}. In order to keep the new system close to the
irreversible system, it is necessary to correctly choose the
parameters so that the previously irreversible reactions remain at
least quasi-irreversible; i.e. the previously neglected reverse reactions
must be slow compared to the direct reactions. As a
consequence, these reactions must have a very negative
$\Delta_rG^0$. This corresponds to big energy differences between
monomers and activated monomers, between activated monomers
and dimers, and between dimers and monomers. The diagram of
Fig.~\ref{fig:energ-level} summarizes the energetic profile of the
system.

A simple source of energy is now explicitly introduced to show the
origin and repartition of fluxes inside the system. In so doing, we
have to take into account an activating agent $X$ that allows the
transformation of monomer into activated monomer. The activation
reaction will thus become $L+X\rightleftharpoons L^*$, essentially
displaced to the right. Additionally, the spontaneous deactivation of
the activated monomer will release a low potential molecule $Y$ as
a waste. This can be done either via a direct deactivation back to
monomer $L^*\rightleftharpoons L+Y$, or an indirect one through
polymerization $L^*+L \rightleftharpoons LL + Y$, these two reactions
being again essentially displaced to the right. A huge difference of energy
must exist between $X$ and $Y$ to guarantee the
quasi-irreversibility of the reactions. The transfer of energy is
globally brought by the transformation of $X$ into $Y$, as a result of the
activation/deactivation process.

Like for the Onsager's triangle, the system can be decomposed into
an internal subsystem (corresponding to the previous system with
explicit energy fluxes) and an exchange subsystem:
{\scriptsize%  
\sublabon{equation}                        %Uncomment for formatted
\begin{eqnarray}
  \boldsymbol{\nu}
%= \begin{bmatrix}                              % Comment for formatted
  &=&                                       % Uncomment for formatted
  \left[                                    % Uncomment for formatted
 \begin{smallmatrix}                        % Uncomment for formatted
    -1        & \cdot & +1    & \cdot & -1    & \cdot & -1    & \cdot & +2    & +1    & +1    & \cdot & \cdot & \cdot & -1 \\
    \cdot     & -1    & \cdot & +1    & \cdot & -1    & \cdot & -1    & \cdot & +1    & +1    & +2    & \cdot & \cdot & +1\\
    +1        & \cdot & -1    & \cdot & -1    & -1    & \cdot & \cdot & \cdot & \cdot & \cdot & \cdot & \cdot & \cdot & \cdot\\
    \cdot     & +1    & \cdot & -1    & \cdot & \cdot & -1    & -1    & \cdot & \cdot & \cdot & \cdot & \cdot & \cdot & \cdot\\
    \cdot     & \cdot & \cdot & \cdot & +1    & \cdot & \cdot & \cdot & -1    & \cdot & \cdot & \cdot & +1    & \cdot & \cdot\\
    \cdot     & \cdot & \cdot & \cdot & \cdot & +1    & \cdot & \cdot & \cdot & -1    & \cdot & \cdot & \cdot & -1    & \cdot\\
    \cdot     & \cdot & \cdot & \cdot & \cdot & \cdot & +1    & \cdot & \cdot & \cdot & -1    & \cdot & -1    & \cdot & \cdot\\
    \cdot     & \cdot & \cdot & \cdot & \cdot & \cdot & \cdot & +1    & \cdot & \cdot & \cdot & -1    & \cdot & +1    & \cdot\\
    -1        & -1    & \cdot & \cdot & \cdot & \cdot & \cdot & \cdot & \cdot & \cdot & \cdot & \cdot & \cdot & \cdot & \cdot\\
    \cdot     & \cdot & +1    & +1    & +1    & +1    & +1    & +1    & \cdot & \cdot & \cdot & \cdot & \cdot & \cdot & \cdot
\end{smallmatrix}                           % Uncomment for formatted
  \right]   \\                                 % Uncomment for formatted
%\end{bmatrix}                                % Comment for formatted
 & \Rightarrow &
  \mathrm{Null}(\boldsymbol{\nu})=
  \bordermatrix{%
       & &\cr
    L  &1&\cdot\cr
    D  &1&\cdot\cr
    L^*&1&1\cr
    D^*&1&1\cr
    LL &2&\cdot\cr
    LD &2&\cdot\cr
    DL &2&\cdot\cr
    DD &2&\cdot\cr
    X  &\cdot&1\cr
    Y  &\cdot&1\cr
  }.
\end{eqnarray}
\sublaboff{equation}                      % Uncomment for formatted
} 
The first vector of the left null space base gives a pool of mass
conservation for $\{L$, $D$, $L^*$, $D^*$, $2LL$, $2LD$, $2DL$,
$2DD\}$, and the second vector for $\{L^*, D^*,X,Y\}$. 
$X$ and $Y$ compounds are the only ones involved in matter exchanges, so
that their concentrations, as well as the concentrations of $L^*$ and
$D^*$ are ruled by the external conditions. The internal subsystem
corresponds to the system of monomers, with the conservation of their
total number: 
\begin{equation}
  c_{tot}=l+d+l^*+d^*+2\cdot ll +2\cdot ld +2\cdot dl+2\cdot dd 
\end{equation}
This property is characteristic of recycled systems. For open-flow
systems, such as the Frank model, there is no internal mass
conservation. All compounds are involved in a linear chain of reactions
connected to external sources (see the Appendix part ``recycled and open-flow
systems'' for more details).

As stated in the original articles\cite{plasson-07,plasson-04}, this
kind of activation can correspond to the activation of amino acids
into NCA (N-carboxyanhydride of $\alpha$-amino acids). Several
activating agents $X$ can be used. In this context,
carbonyldiimidazole (CDI) is a commonly used compound\cite{wen-01,
  *hitz-03} that may have been formed in prebiotic
conditions\cite{saladino-06}. Other prebiotically relevant agents are
also often cited in the literature, like activation by isocyanate and
nitrogen oxides\cite{taillades-99}, or carbon monoxides in conjunction
with sulfide compounds\cite{huber-98, *leman-04,
  *wachtershauser-07}. When the activation leads to NCA formation, the
waste compound $Y$ resulting from the spontaneous hydrolysis of NCA is
CO$_2$. As required, all these $X$ compounds are very reactive, while
$Y$ is a very stable compound\cite{balance}. One could additionally
focus on the phenomenon that leads to the presence and maintenance of
$X$ compounds on prebiotic Earth (like stable nonequilibrium
concentrations of nitrogen oxides or carbon monoxide in the
atmosphere\cite{mancinelli-88, *kasting-90}, or the production of
sulfides by geochemical processes\cite{cody-04}). As long as these
external phenomena are present, a chemical source of energy would be
available, keeping the chemical system of amino acids in an active
nonequilibrium state.

\subsection*{Thermodynamic and Kinetic Parameters}

The relations will be written as a function of the
parameters:
\sublabon{equation}
 \begin{eqnarray}
  K_{act}&=&\frac{K_{f,L^*}}{K_{f,L}} \ll 1     ,\\
  K_{dim}&=&\frac{K_{f,LL}}{K_{f,L}^2} \ll 1     ,\\
  K_{epi}&=&\frac{K_{f,LD}}{K_{f,LL}}=\gamma     .
\end{eqnarray} 
\sublaboff{equation}%
$K_{act}$ depends on the difference of energy between $L^*$ and $L$
(or, equivalently, between $D^*$ and $D$).  $K_{dim}$ depends on the
difference of energy between $L$ and $LL$ (or $D$ and $DD$). And
$K_{epi}$, in turn, depends on the difference of energy between $LD$
and $DD$ (or $DL$ and $LL$). Thus, the situation is totally
symmetrical with respect to interchanges between $L$ and $D$ above.

\begin{table*}[btp]
  \centering
  \caption{Kinetic and thermodynamic parameters of the reactions of
    the APED system. Kinetic parameters of reactions involving $X$ or
    $Y$ compounds are apparent kinetic rates, assuming constant
    concentration of $x$ and $y$. The exact symmetrical of each of these
    reactions is also included in the whole network, with
    exactly the same parameters.}
  \label{tab:kin}
  \begin{tabular}{c|cccc}
    Reaction& $k_j$ & $K_j$ & $k_{-j}$ & $\frac{\sigma_j}{R}$\\
    \hline
    Activation:\\
    $X + L 
    %\xrightleftharpoons[k_{-A}]{k_A}
    \underset{k_{-A}}{\stackrel{k_A}{\rightleftharpoons}}
    L^*$ & $k_A$ &
    $\frac{K_{act}}{K_{\mathrm{f},x}}$ & $\frac{k_A}{K_{act}V_x}$ &
    $k_A( l- \frac{1}{K_{act}V_x} \cdot
    l^*)\ln\frac{K_{act}V_xl}{l^*}$\\ 
    Deactivation:\\
    $L^* 
    % \xrightleftharpoons[k_{-H}]{k_H} 
    \underset{k_{-H}}{\stackrel{k_H}{\rightleftharpoons}}
    L+Y$ & $k_H$& 
    $\frac{K_{\mathrm{f},y}}{K_{act}}$ & $ k_HK_{act}V_y$ & 
    $k_H( l^*-K_{act}V_y l)\ln\frac{l^*}{K_{act}V_yl}$\\ 
    Polymerization:\\
    $ L^* + L  
    %\xrightleftharpoons[k_{-P,1}]{k_{P,1}} 
    \underset{k_{-P,1}}{\stackrel{k_{P,1}}{\rightleftharpoons}}
    LL + Y$ & $k_P$& 
    $\frac{K_{dim}K_{\mathrm{f},y}}{K_{act}}$ & $ k_P\frac{K_{act}V_y}{K_{dim}}$ & 
    $k_P( l \cdot l^*- \frac{K_{act}V_y}{K_{dim}} ll)\ln\frac{K_{dim}l\cdot l^*}{K_{act}V_yll}$\\ 
     $ L^* + D  
     %\xrightleftharpoons[k_{-P,2}]{k_{P,2}} 
     \underset{k_{-P,2}}{\stackrel{k_{P,2}}{\rightleftharpoons}}
     LD + Y$ &$\alpha k_P$& 
    $\gamma\frac{K_{dim}K_{\mathrm{f},y}}{K_{act}}$ & 
    $k_P \frac{\alpha}{\gamma} \frac{K_{act}V_y}{K_{dim}}$ & 
    $\alpha k_P( d \cdot l^*-\frac{K_{act}V_y}{\gamma K_{dim}} ld)\ln\frac{\gamma K_{dim}d\cdot l^*}{K_{act}V_yld}$\\ 
    Depolymerization:\\
    $LL  
    %\xrightleftharpoons[k_{-D,1}]{k_{D,1}} 
    \underset{k_{-D,1}}{\stackrel{k_{D,1}}{\rightleftharpoons}}
    L + L$ & $k_D$& 
    $\frac{1}{K_{dim}}$ & $ k_DK_{dim}$ & 
    $k_D(ll - K_{dim} \cdot l^2)\ln\frac{ll}{K_{dim}l^2}$\\ 
    $  LD  
    % \xrightleftharpoons[k_{-D,2}]{k_{D,2}} 
    \underset{k_{-D,2}}{\stackrel{k_{D,2}}{\rightleftharpoons}}
    L + D$ & $\beta k_D$& 
    $\frac{1}{\gamma K_{dim}}$ & $ \beta\gamma k_DK_{dim}$ & 
    $\beta k_D( ld - \gamma K_{dim} \cdot l \cdot
    d)\ln\frac{ld}{\gamma K_{dim}l\cdot d}$\\ 
    Epimerization:\\
    $DL
    %\xrightleftharpoons[k_{-E}]{k_E}
    \underset{k_{-E}}{\stackrel{k_E}{\rightleftharpoons}}
    LL$ & $k_E$ &
    $\frac{1}{\gamma}$& $\gamma k_E$ &$k_E(dl-\gamma \cdot
    ll)\ln\frac{dl}{\gamma ll}$ \\
    Racemization:\\
    $  L  
    %\xrightleftharpoons[k_R]{k_R} 
    \underset{k_{-R}}{\stackrel{k_R}{\rightleftharpoons}}
    D$ & $k_R$ &
    $1$& $k_R$ &$k_R(l-d)\ln\frac{l}{d}$ \\
   \hline
  \end{tabular}
\end{table*}

The ensemble of kinetic parameters relative to all the chemical
reactions involved in the system is given in Tab.~\ref{tab:kin}. 
This way, the system can be totally characterized, while remaining
perfectly compatible with thermodynamic and kinetic relations, by the
following independent parameters:
\begin{itemize}
\item Thermodynamic: $K_{act}$, $K_{dim}$, $\gamma$
\item Kinetic: $k_A$, $k_H$, $k_P$, $k_D$, $k_E$, $k_R$, $\alpha$, $\beta$
\item External conditions: $V_x$, $V_y$
\item Internal conditions: $l$, $d$, $l^*$, $d^*$, $ll$, $ld$, $dl$
  and $dd$
\end{itemize}
By choosing the parameters so that all the back reactions are
negligible comparing to the direct reactions, the system described in
the previous study, not taking into account microreversible
reactions\cite{plasson-04}, would be a correct approximation of the
complete reversible framework described here. We thus need, in
accordance with the first assumption of energy levels:
%\sublabon{equation}
 \begin{eqnarray}
  V_x&\gg&\frac{1}{K_{act}}    ,\\
  K_{act}\cdot V_y &\ll&K_{dim} \ll 1   .
\end{eqnarray} 
%\sublaboff{equation}%
This corresponds to a big difference of energy between mono\-mers
and di\-mers, a yet larger difference between monomers and
activated monomers, and a very high potential of the activating
agent $X$. Assuming that $x$ and $y$ are constant --- maintained by
external phenomena, or present in a large excess --- the reactions
involving $X$ and $Y$ compounds correspond to pseudo first
order reactions in monomers or dimers. Mathematically, the system is
equivalent to a closed system in monomer derivatives, maintained in
a nonequilibrium state by a continuous flux of chemical energy
brought by the spontaneous reaction from $X$ to $Y$, these compounds
being maintained by other external processes.

The system is subject to a continuous flux of chemicals:
\begin{eqnarray}
  \sigma^\mathrm{e}&=&R
  \left(
    \varphi_X^e\ln V_x+\varphi_Y^e\ln V_Y
  \right)   .
\end{eqnarray} 
In the steady state, the incoming flux of $X$ is equal to the outgoing
flux of $Y$, so that:
 \begin{eqnarray}
  \varphi^e_X&=&\varphi^e_y=\varphi^e   .
\end{eqnarray}
It will be compensated by  the consumption rate of $X$:
\begin{eqnarray}
  v_X&=&0   ,\\
  \varphi^e&=&k_A(l+d)-k_{-A}(l^*+d^*)   .
\end{eqnarray} 
As a consequence, the steady-state flux of entropy of exchange is:
\begin{eqnarray}
  \sigma^\mathrm{e}&=&Rk_A\left(
    \left(l+d\right)-\frac{1}{K_{act}V_x}\left(l^*+d^*\right)
  \right)\ln\frac{V_x}{V_y}  \label{eq:incom-ent} .
\end{eqnarray}

\subsection*{Analysis of the System}

Simulations were performed\cite{xppaut} with $c_{tot}=2$~M , $k_A=1~\mbox{s}^{-1}$,
$k_H=1~\mbox{s}^{-1}$, $k_P=1~\mbox{s}^{-1}\mbox{.M}^{-1}$,
$k_D=1~\mbox{s}^{-1}$, $k_E=1~\mbox{s}^{-1}$,
$k_R=0.0001~\mbox{s}^{-1}$, $\alpha=0.1$, $\beta=0.1$, $\gamma=0.1$,
$K_{act}=10^{-6}$, $K_{dim}=10^{-3}~\mbox{M}^{-1}$,
$K_{f,x}=10^{-9}~\mbox{M}$ and $V_y=1$. The parameter $x$ was kept as
a variable, allowing to tune the distance from the equilibrium.

\begin{figure}[hb]
\centering
  \includegraphics[width=6cm,clip]{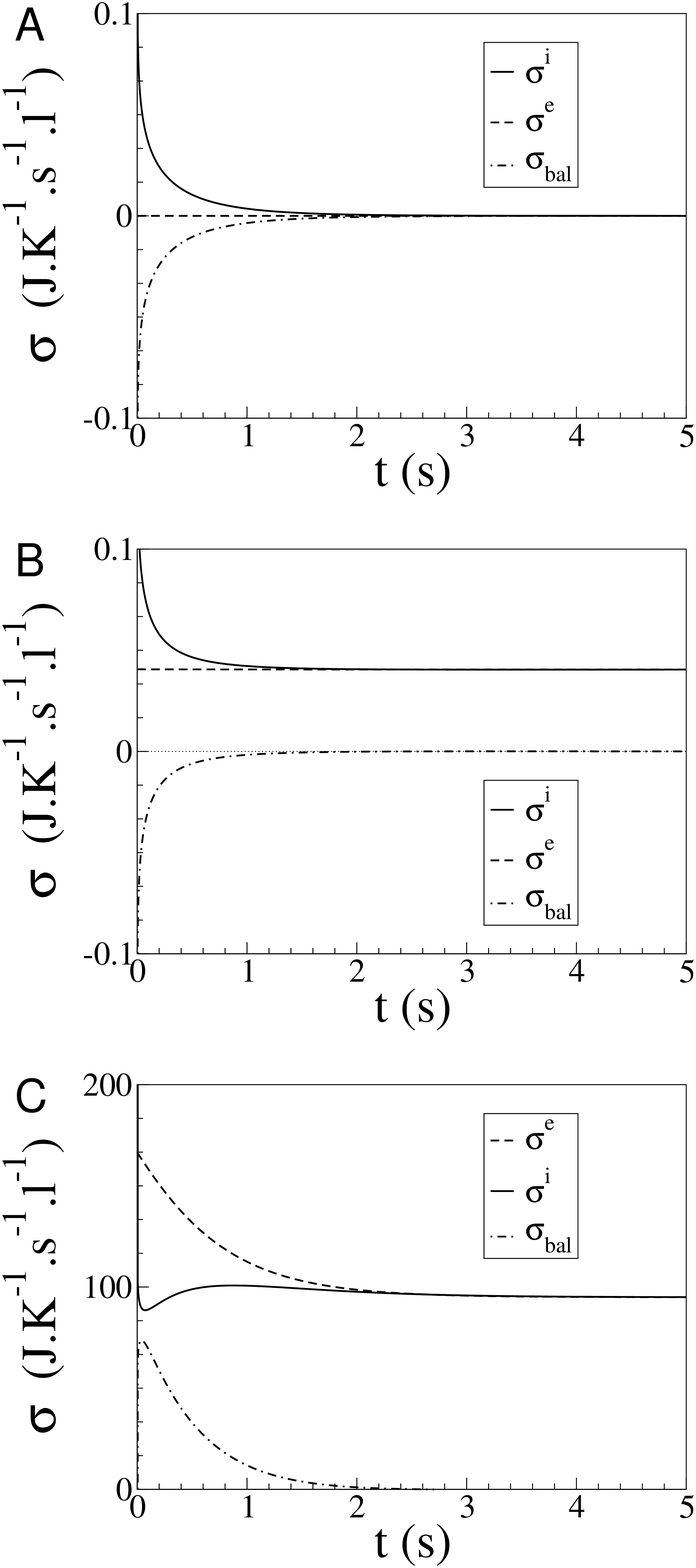}
  \caption{Entropy production of APED systems during the early
    evolution. A, $x=0$~M; B, $x=2.10^{-4}$~M; C, $x=0.5$~M. See
    the text for the other parameter values.}
\label{fig:early-evol}
\end{figure}

\subsubsection*{Early evolution.}
Fig.~\ref{fig:early-evol} represents the early evolution of the system,
reaching quickly the racemic steady state, from an initial state
composed exclusively by a quasi-racemic mixture of monomers
(initial enantiomeric excess $10^{-5}$). In this short period, the
system reaches a racemic steady state, i.e. a steady state that is
issued from the thermodynamic equilibrium branch. As observed in the
nonequilibrium Onsager's triangle, different behaviors are
characteristic of equilibrium close-to-equilibrium and
far-from-equilibrium conditions: 
\begin{description}
\item[Fig.~\ref{fig:early-evol}A]: there is no energy source, the
  system reaches the equilibrium state. The entropy production
  decreases towards zero. This system dissipates its initial excess of
  energy before reaching the isolated equilibrium state.
\item[Fig.~\ref{fig:early-evol}B]: there is a weak energy source so
  that the system remains close to the equilibrium state. The
  evolution is very similar to the preceding one, except that the
  entropy production decreases until a minimal value. The
  system continuously dissipates the incoming energy. Similarly to the
  previous system, the system initially dissipates its excess of
  energy. 
\item[Fig.~\ref{fig:early-evol}C]: there is a strong energy source, the
  system is far from the equilibrium state. The balance of entropy is
  now positive: the system stocks energy --- rather than
  dissipating it as previously observed --- by acquiring high
  concentration of high potential compounds.
\end{description}

\begin{figure}[ht]
  \centering
  \includegraphics[width=6cm,clip]{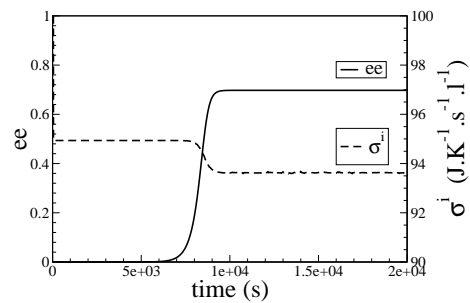}
  \caption{Decrease of the entropy production in an APED system during the transition
    from an unstable racemic steady state towards a stable non-racemic
    steady state, in the case $x=0.5$~M.}
  \label{fig:bifur-entr}
\end{figure}

Following a longer period, we observe a transition of state for some
values of $x$ (see Fig.~\ref{fig:bifur-entr}): the enantiomeric excess
abruptly changes to a non-zero value. The system switches from the
unstable racemic branch towards the stable non-racemic branch. This
transition is followed by a decrease of entropy production. A similar
dynamical behavior is observed in the Frank model\scite{kondepudi-08}
while the chemistry underlying both models is rather different. If
variations of entropy production are typically observed during
transition between states of different stability, as in
autocatalytic\scite{yoshida-90} or oscillating\scite{dutt-90} systems,
the sign of this variation is not necessarily negative.

\subsubsection*{Complete Bifurcation Pattern.}

\begin{figure*}[ht]
  \centering
    \includegraphics[width=12cm,clip]{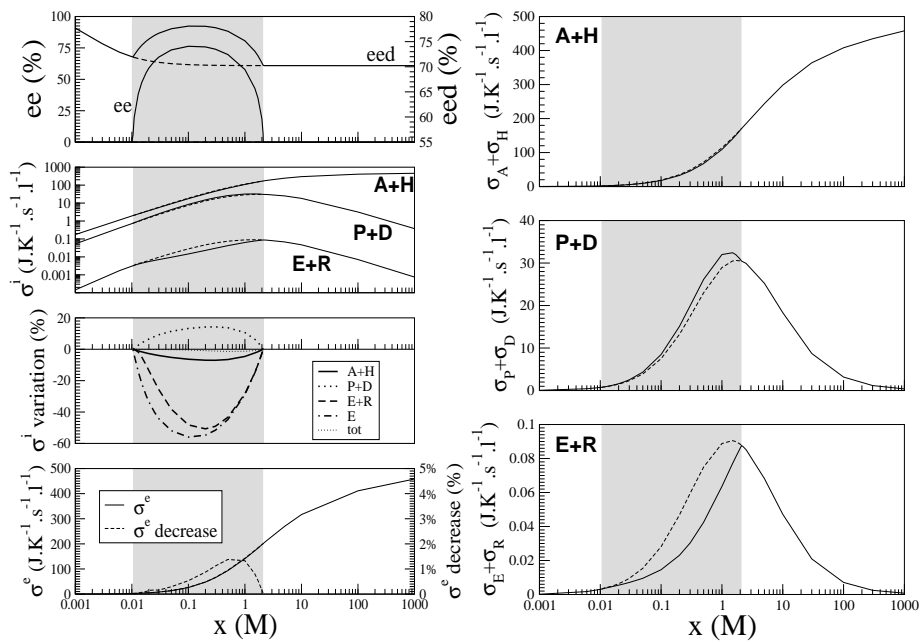}
    \caption{Enantiomeric excess ($ee$), diastereoisomeric
      excess of dimers
      ($eed=\frac{([LL]+[LD])-([LD]+[DL])}{[LL]+[DD]+[LD]+[DL]}$) and
      entropy production in the steady states as a function of the
      incoming flux of energy in APED systems (see text for
      parameters). Solid curves correspond to values for stable steady
      states, and dotted curves to unstable steady states. A, H, P, D,
      E and R stand for activation, deactivation, polymerization,
      depolymerization, epimerization and racemization,
      respectively. The gray area corresponds to the range where
      non-racemic states are stable.}
\label{fig:bif-entr}
\end{figure*}

The same system was studied in detail for different values of $x$,
ranging from $10^{-3}$~M to $10^3$~M. The particulars of the different
time courses obtained are reported in Fig.~\ref{fig:bif-entr}. A
different bifurcation pattern than the one usually found is now
observed:
\begin{itemize}
\item For low values of $x$, the system remains racemic. As
  expected, this corresponds to a state where the system remains close
  to the equilibrium since not enough energy is available.
\item Then, starting from a critical value of $x=0.0105$~M, a bifurcation point
  is observed. The system switches to a non racemic branch, a
  far-from-equilibrium state that becomes more stable than the
  close-to-equilibrium one. This corresponds to the expected behavior
  happening when enough energy is available.
\item However, for the value $x=0.15$~M, the enantiomeric excess
  actually reaches a maximum, and then decreases back to a second
  bifurcation point at $x=2.11$~M. The system then switches back to
  the racemic branch as the only remaining stable steady state. This
  behavior can actually be also observed in the complete Frank
  model\cite{cruz-08} or in open-flow autocatalytic
  systems\scite{yoshida-90}. 
\end{itemize}

\subsubsection*{Variation of Entropy Production during the Transitions.}

A monotonous increase of the incoming energy $T\sigma^\mathrm{e}$ as a
function of $x$ is observed. In the non racemic stable state --- when
it exists --- the system consumes up to $2\%$ less energy than on the
racemic branch. This corresponds to the decrease of the total entropy
production during the transition between steady-states.  This decrease
is mostly due to a decrease of entropy production by
activation/deactivation reactions, and by the epimerization
reactions. These decreases are small in absolute value, but the
relative variation of entropy production is very important for the
epimerization reactions (up to $60\%$).

We thus observe a small change on the global entropy production of the
system, that allows a large local change centered in the
epimerization reactions --- the engine of the deracemization process,
as they are the only reactions able to change the $L/D$ balance in a
system in which the total number of monomers is constant --- and a
local increase of the entropy production by the
polymerization\slash{}depolymerization processes --- that is, a
redirection a larger part of the energy fluxes towards the
autocatalytic loops. Globally, the switch to the far-from-equilibrium
branch allows the use of the incoming energy to both produce more
dimers and to spontaneously favour the stable non-racemic state.

\subsubsection*{Evolution of the Entropy Production with the Energy Flux.}

The global entropy production follows the global energy consumption
and thus also monotonously increases as a function of $x$. For all
systems, the energy is mostly dissipated by the activation and
deactivation reactions. More interestingly, the entropies produced by
the polymerization, depolymerization and epimerization reactions are
actually increasing when stable non-racemic steady states exist, but
decrease to zero after the second bifurcation point. That is, after
the second bifurcation point, the
activation\slash{}polymerization\slash{}epimerization\slash{}depolymerization
loops disappear, giving place to only
activation\slash{}deactivation loops.

\subsubsection*{Interpretation.}

\begin{figure}[hb]                                   %Uncomment for formatted
%\begin{figure}[hbt]                                  % Comment for formatted
  \centering
  \includegraphics[width=8cm]{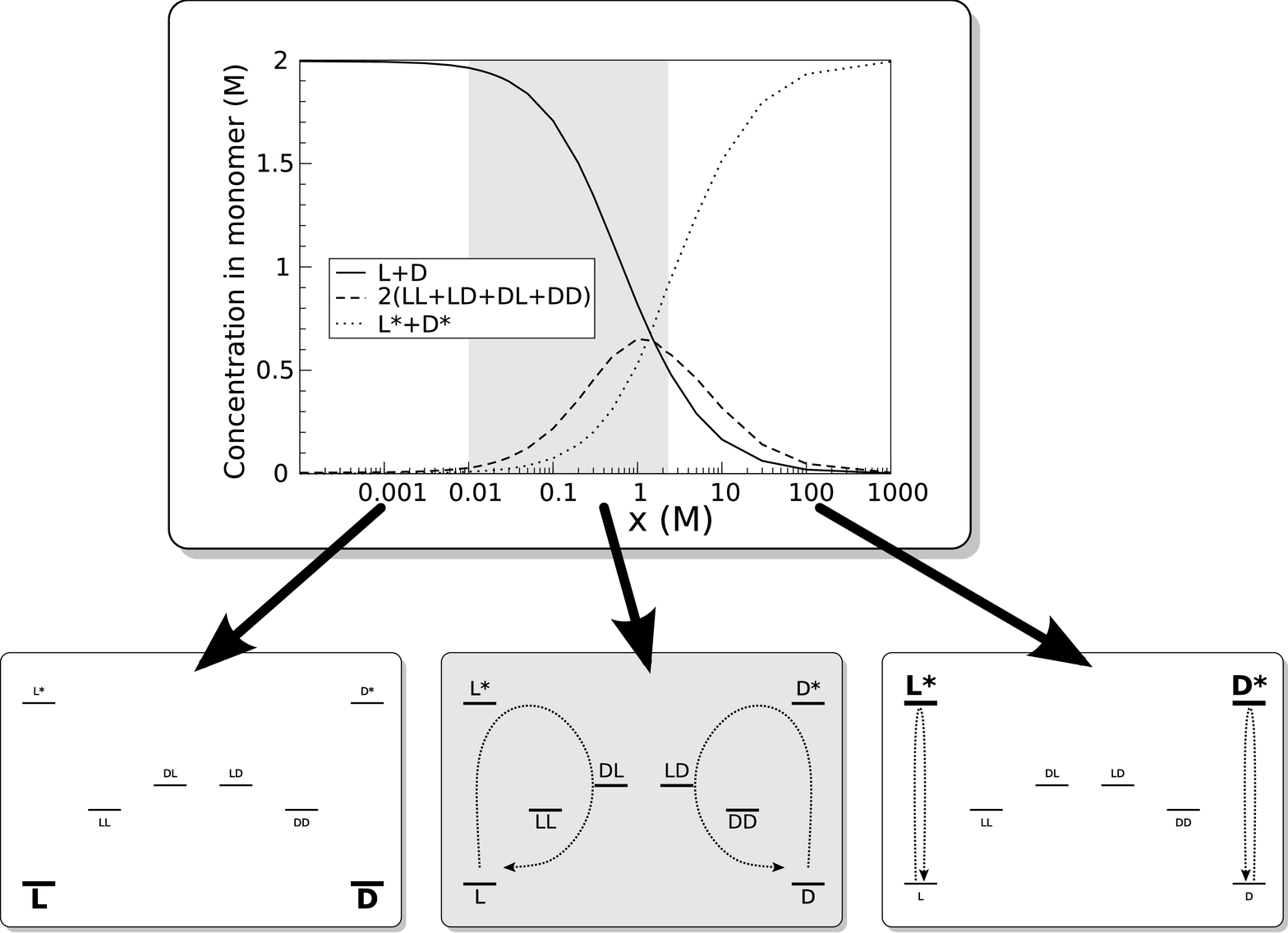}              %Uncomment for formatted
  \caption{Steady-state composition in APED systems for different
    energy fluxes. Top figure: steady state concentrations of free,
    polymerized and activated monomers as a function of the
    concentration of the activating agent $X$. Bottom figures:
    schematic representation of major cycle fluxes before the first
    bifurcation point, between the two bifurcation points, and after
    the second bifurcation point; the letter sizes represent the
    relative quantity of compounds; the arrows represent the major cycle
    fluxes. The gray area corresponds to the range where non-racemic
    states are stable.}
  \label{fig:rep}
\end{figure}

When more and more energy is introduced into the system, more and
more compounds are maintained at high level of energy. As the key
compounds (i.e.: the dimers, since they allow differences of energy
between the diastereoisomers) are at intermediate energy levels,
the deracemization will only occur for an interval of
energy flux. Not enough energy can not produce enough dimers while
too much energy destroys the dimers in favor of activated monomers (see
Fig.~\ref{fig:rep}).

Before the first bifurcation, the system is very close to
equilibrium, the most populated states being the lower energy states
(i.e. the free monomers). There is just not enough energy to be
directed towards the autocatalytic cycles.

After the second bifurcation point, the situation is reversed. The
high flux of energy is almost only directed to the
activation\slash{}deactivation loops. The most populated states are
the highest energy states. This is characterized by a saturation
phenomenon with most of the monomers being maintained in the activated
form, rather than in dimers or free monomers. Just a small amount of
energy is directed towards the autocatalytic loops. This phenomenon
can be compared to what is observed in open-flow Frank-like
systems\cite{cruz-08}. When matter fluxes are too important, the system
remains racemic: the high energy incoming compounds have no time to
react before being flushed outside the system. Similarly to what is
observed here, the excess of energy flux (due to matter exchanges)
maintains high concentration of high energy compounds, preventing them
to react in autocatalytic reactions.

It is only between the two bifurcation points that the autocatalytic
loops can be efficiently performed. A substantial quantity of
dimers is maintained, so that the epimerization reaction can
effectively take place. The energy is thus efficiently used to
generate a non-racemic steady state.

\section*{Conclusion}

The full thermodynamic study of the APED system emphasizes the great
importance of the energy transfers. Energy must flow into the system
in order to stabilize the non racemic state. But just the consumption
of energy is not sufficient: energy must be \emph{efficiently}
used. It must be directed towards the interesting elements of the
reaction network, that is the autocatalytic loops, able to induce
positive feedbacks inside the system. This point has to be focused on
when designing chemical experimental systems that can present
such non-linear behaviors.

This redirection of some part of the free energy available in the
environment (here, the difference of chemical potentials maintained
between $X$ and $Y$ compounds) towards some internal ``mechanisms''
(here, the chiral autocatalytic cycles) is one of the key conditions
allowing the emergence of nonequilibrium properties (here, a
non-racemic steady state). Studies of the origin of life have to take
into account such type of ``energy focus''\cite{bergareche-99,
  *ruiz-mirazo-04}\textsuperscript{,}\scite{morowitz-68}\textsuperscript{,}\scite{morowitz-92}.
Real chemical systems are generally away from the equilibrium state as
they continuously communicate with their surrounding. Several abiotic
systems can be found in very different environments (interstellar
space\cite{bourlot-95}, planet crust\cite{cody-04} or
atmosphere\cite{mancinelli-88, *kasting-90}) and diverse forms of
available free energy can be found in these environments: by means of
these external energy sources, nonequilibrium concentrations of
activated compounds can be maintained.

In this context, the challenge is to better understand how this
available energy can be passed down to other chemical systems and used
so that it leads to self-organization, in a similar way as it is
performed by metabolisms. This corresponds to understanding the
emergence of protometabolic systems\cite{duve-07}. When a chemical
system is connected to chemical sources of energy, internal reaction
loops are possibly maintained, potentially leading to the emergence of
autocatalytic behaviors\scite{morowitz-07}. Such nonequilibrium
chemical systems are prone to self-organization and
self-maintenance\cite{eigen-71, farmer-86}. Thus, the transition of
such protometabolic systems towards replicative systems can be seen as
a process of spontaneous self-organization under energetic
pressures. Rather than opposing self-organization and natural
selection\cite{fernando-07}, it may help to understand how natural
selection emerges from self-organization\cite{hoelzer-06}. Avoiding
the vitalist idea of biochemical systems behaving in a fundamentally
different way than abiotic chemical systems, it leads to the
description of general chemical systems under similar principles.

As a consequence, if a given system permit the continuous activation
of amino acids with a sufficient energetic efficiency\cite{wen-01, *hitz-03},
polymerization/depolymerization cycles will naturally be
established. This would correspond to a very simple self-organized
protometabolism, able to consume energy, to generate some complexity
by synthesizing and maintaining a dynamic pool of oligopeptides, and
potentially being able to evolve towards a stable non racemic
state. If such scenario could have been realized, it would imply
mirror symmetry breaking to be one natural property of self-organizing
protometabolisms. Of course, strong precautions must be taken when
dealing with such complex chemical systems, as emphasized by
Orgel\cite{orgel-08}. The robustness, stability, and chemical
relevance of this scenario must be thoroughly tested and examined.

{\small
  \subsubsection*{Acknowledgment:} K. Asakura, R. Pascal, D.K. Kondepudi
  and A. Brandenburg are gratefully acknowledged for comments and
  discussions.

\section*{Appendix}

\subsubsection*{Microreversibility and Detailed Balancing.}

It is important to understand that chemical reactions are reversible
at the microscopic level, and irreversible at the macroscopic
level. The microscopic reversibility (or microreversibility) is a
consequence of the time-reversal symmetry of the fundamental laws of
physics: for a given molecule subjected to a chemical transformation,
the inverse transformation can be obtained by following the exact
pathway but in the opposite direction. The detailed balancing is a
macroscopic property that can be deduced from microreversibility: when
the system is at thermodynamic equilibrium, reaction rates in both
directions cancel for each reaction (implying the relationship given
in Eq.~\ref{eq:detailed-bal}). This transition from a microscopic to a
macroscopic description --- involving statistic treatment and the loss
of information about individual molecule behaviors --- explains how as
chemical system becomes irreversible at the macroscopic level, in the
sense that chemical reactions always evolve spontaneously from a
nonequilibrium state towards an equilibrium state, never in the
opposite direction. More details on this subject can be found in the
article of Boyd\cite{boyd-77}.

In that context, there is thus no detailed balancing in a
nonequilibrium system, in the sense that not all the individual fluxes
have to be cancelled: there can be the onset of chemical fluxes inside
the system\cite{klein-55}. However, this does not imply that there is
a breakdown of time-reversal symmetry at the microscopic level (as it
may be the case in some specific systems\cite{barron-87, *barron-94}).
In a nonequilibrium system, there can be chemical fluxes, but kinetic
constants remain linked through the thermodynamic equilibrium
constants.

\subsubsection*{Kinetic Relationships.}

The system exchanges matter with its surrounding. The variation in
concentration due to external exchanges, for each incoming or outgoing
compound, is:
\begin{eqnarray}
  \ddt[e]{x_i}&=&\varphi_i^\mathrm{e} .\label{eq:def-flux-start}
\end{eqnarray}
Note that $\varphi^\mathrm{e}_i$ can be positive or negative, depending on whether the
component comes into the system or leaves it. Internal transformations
occur through the chemical processes. The variation in concentration due
to the internal transformations is:
%\sublabon{equation}
\begin{eqnarray}
  \varphi_j^+&=&k_j\prod_{i=1}^n x_i^{\nu^-_{i,j}}   ,\\ 
  \varphi_j^-&=&k_{-j}\prod_{i=1}^n x_i^{\nu^+_{i,j}}    ,\\ 
  \varphi_j&=& \varphi_j^+-\varphi_j^- \label{eq:kin-i1}   ,\\ 
  \ddt[i]{x_i}  &=&  \sum_{j=1}^r \nu_{i,j} \varphi_j  .
\end{eqnarray} 
%\sublaboff{equation}%
The total variation in the concentration of each chemical species is
thus given by:
\begin{eqnarray}
  v_i &=&\ddt{x_i}    ,\\
  &=&\ddt[e]{x_i}+\ddt[i]{x_i}\label{eq:def-flux-end} .
\end{eqnarray}

\subsubsection*{Microreversibility.}

At equilibrium, the microreversibility implies the detailed balancing
of the reaction (i.e. the fluxes are canceling in both directions of
each reaction):
\begin{eqnarray}
  \label{eq:def-gamma-start}
  \varphi_j^+&=&\varphi_j^-       ,\\ 
  \frac{k_j}{k_{-j}}&=&\prod_{i=1}^n x_{i,eq}^{\nu_{i,j}}    ,\\
  &=&K_j \label{eq:reac-kin} .
\end{eqnarray}
The equilibrium constant can be expressed as a function of the
standard constants of formation of the compounds involved in the reaction:
\begin{eqnarray}
  K_j&=&\prod_{i=1}^n K_{\mathrm{f},i}^{\nu_{i,j}} .\label{eq:reac-form}
\end{eqnarray} 
It is then possible to express a relation between the parameters of
the direct reaction with the parameters of the indirect reaction:
\begin{eqnarray}
  k_{-j}\displaystyle\prod_{i=1}^n K_{\mathrm{f},i}^{\nu^+_{i,j}}&=&
  k_j\displaystyle\prod_{i=1}^n K_{\mathrm{f},i}^{\nu^-_{i,j}}   ,\\
  &=&\Gamma_j   .\label{eq:def-gamma-end}
\end{eqnarray}
This leads to the expression of $\Gamma_j$, that is a characteristic
parameter, combining kinetic and thermodynamic properties of the
microreversible reaction.

\subsubsection*{Relation fluxes-potential.}

\begin{eqnarray}                                                       
  \varphi_j&=&k_j\displaystyle\prod_{i=1}^n                          
  K_{\mathrm{f},i}^{\nu^-_{i,j}}\cdot \frac{\displaystyle\prod_{i=1}^n  
    x_i^{\nu^-_{i,j}}} {\displaystyle\prod_{i=1}^n                      
    K_{\mathrm{f},i}^{\nu^-_{i,j}}}                                     
  -k_{-j}\displaystyle\prod_{i=1}^n                                    
  K_{\mathrm{f},i}^{\nu^+_{i,j}}\cdot \frac{\displaystyle\prod_{i=1}^n  
    x_i^{\nu^+_{i,j}}}                                                  
  {\displaystyle\prod_{i=1}^n K_{\mathrm{f},i}^{\nu^+_{i,j}}} \label{eq:demo-ohm}\\    
   &=&\Gamma_j
  \left(
    \prod_{i=1}^n V_i^{\nu^-_{i,j}}
    -
    \prod_{i=1}^n V_i^{\nu^+_{i,j}}
  \right)   .  
\end{eqnarray}
The parameters relative to the reaction ($\frac{\varphi_j}{\Gamma_j}$)
are separated from the parameters relative to reactants ($\prod
V_i^{\nu^-_{i,j}}$) and from the parameters relative to the products
($\prod V_i^{\nu^+_{i,j}}$). This equation is analogous to an ``Ohm's
law'', in which the reaction flux ($\varphi_j$), as an analogous to an
electrical intensity, becomes a function of the potential between
reactants and products ($V_i$ parameters). Similar laws are given in
the literature, but the relationship is sometimes given as a function
of chemical potentials $\mu_i$ rather than the $V_i$ parameters, in
which case this law becomes true only close to
equilibrium\cite{qian-05}. The relation of Eq.~\ref{eq:demo-ohm} is true even in
nonequilibrium states, and is rather similar to the description of
Peusner \emph{et al.}\cite{peusner-85}, except that the $V_i$
potentials are calculated relatively to the standard state --- thus
the introduction of the $K_{\mathrm{f},i}$ --- rather than relatively
to one arbitrary node.  The advantage of this description is to be
totally symmetric for all compounds, not depending on any numbering,
and is thus more general. 

\subsubsection*{Entropic Analysis.}

The system will continuously dissipate energy, through the chemical
reactions.  The entropy production by unit of time $\sigma_j$, per each
reaction taking place in the system, is equal to\cite{prigogine-62}:
\begin{eqnarray}
  \sigma_j &=&\dpdt[i]{S_j}{P,T}           ,\\
  &=&J_\mathrm{ch}X_\mathrm{ch}, \quad \mathrm{with} \quad
  J_\mathrm{ch}=\varphi_j, \quad X_\mathrm{ch}=\frac{\mathcal{A}}{T}
   ,\\
  &=& \varphi_j \cdot \displaystyle\frac{-\sum_{i=1}^n \nu_i\mu_i}{T}
  ,\\
  &=&-R\varphi_j\ln
  \left(
    \prod_{i=1}^n V_i^{\nu_{i,j}}
  \right)\label{eq:entrop-prod-ann}   ,\\
  &=&R(\varphi_j^+-\varphi_j^-)\ln\frac{\varphi_j^+}{\varphi_j^-} .
\end{eqnarray} 
It can be noted that whatever the sign of $\varphi_j$, in accordance with
the second principle, $\sigma_j$ is always positive (zero in the limit
case $\varphi_j=0$).

The total production of entropy by all internal transformations is the
resultant of all these chemical processes, that is:
\begin{eqnarray}
  \sigma^\mathrm{i}&=&\sum_{j=1}^r \sigma_j   .
\end{eqnarray}

At the same time, the system exchanges entropy through
matter exchange with the surrounding:
%\sublabon{equation}
\begin{eqnarray}
  \sigma^\mathrm{e} &=&\frac{1}{T} \sum_{i=1}^n  \ddt[e]{x_i} \mu_i   ,\\
  &=&R\sum_{i=1}^n  \varphi_i^\mathrm{e}\ln V_i\label{eq:entrop-exch-ann} .
\end{eqnarray} 
%\sublaboff{equation}%

\subsubsection*{Recycled and Open-Flow Systems.}

An open-flow system is characterized by the absence of internally
conserved compounds. All of them are linked to matter exchanges, so
that there cannot be any mass conservation rule. The following simple
Frank model is taken as an example:
\sublabon{equation}
\begin{eqnarray}
    &
    \stackrel{\varphi_A^\mathrm{e}}{\rightarrow}
    & A \label{eq:fluxa}\\
    A + L &
    \underset{k_{-1}}{\stackrel{k_1}{\rightleftharpoons}}
    & L + L\\
    A + D &
    \underset{k_{-1}}{\stackrel{k_1}{\rightleftharpoons}}
    & D + D\\
    L + D &
    \underset{k_{-2}}{\stackrel{k_2}{\rightleftharpoons}}
    & P \\
    P & 
    \stackrel{\varphi_P^\mathrm{e}}{\rightarrow} \label{eq:fluxp}   
\end{eqnarray}
\sublaboff{equation}
The conserved moieties of this system are:
{\footnotesize%
\begin{eqnarray}
  \boldsymbol{\nu} =\begin{bmatrix}
    -1          & -1                 & \cdot  \\
    +1          & \cdot              & -1     \\
     \cdot      & +1                 & -1     \\
     \cdot      &  \cdot             & +1     \\
  \end{bmatrix} & \Longrightarrow    &
  \mathrm{Null}(\boldsymbol{\nu})=
  \bordermatrix{%
     &\cr
    A&1\cr
    L&1\cr
    D&1\cr
    P&2
  }.
\end{eqnarray}}
All the compounds are engaged in a linear succession of
reactions connected to exchange fluxes.

In order to recycle this system, the products must be activated back
to the reactants, that is to replace the fluxes of Eq.~\ref{eq:fluxa}
and Eq.~\ref{eq:fluxp} by:
\sublabon{equation}
\begin{eqnarray}
    P + X &
    \underset{k_{-3}}{\stackrel{k_3}{\rightleftharpoons}}
    & 2A + Y\\
    &
    \stackrel{\varphi_X^\mathrm{e}}{\rightarrow}
    & X\\
    Y & 
    \stackrel{\varphi_Y^\mathrm{e}}{\rightarrow}
\end{eqnarray}
\sublaboff{equation}
Which leads to:
{\footnotesize%
\begin{eqnarray}
  \boldsymbol{\nu} =\begin{bmatrix}
    -1          & -1                 & \cdot  & +2\\
    +1          & \cdot              & -1     & \cdot\\
     \cdot      & +1                 & -1     & \cdot\\
     \cdot      &  \cdot             & +1     & -1\\
     \cdot       & \cdot            & \cdot & -1 \\
     \cdot       & \cdot            & \cdot & +1  \\
  \end{bmatrix} & \Longrightarrow    &
  \mathrm{Null}(\boldsymbol{\nu})=
  \bordermatrix{%
     &\cr
    A&1&\cdot\cr
    L&1&\cdot\cr
    D&1&\cdot\cr
    P&2&\cdot\cr
    X&\cdot&1\crcr
    Y&\cdot&1
  }   .
\end{eqnarray}}
This now turns out to be an internal subsystem with mass conservation of
$a+l+d+2p$, connected to the flux of matter $\{X,Y\}$. However, this
system is still not a recycled system of chiral subunits, as there is
no mass conservation of $l+d$, except in the case where the $A$ and
$P$ compounds can be neglected compared to $L$ and $D$. In this latter
case, the process of destruction/construction of chiral subunits
becomes equivalent to an activated racemization process.

\bibliography{biblio}

\begin{mcitethebibliography}{10}
\bibitem{notations}
Following the IUPAC definitions\scite{naught-97}, and in order to avoid
  confusion with the different quantities, the notion of chemical
  fluxes\scite{iupac-chemflux} will be used for the reactions (noted
  $\varphi$), while the notion of rate of appearance\scite{iupac-rateapp} will
  be used for the compounds (noted $v$). A superscript `e' indicates a
  parameter relative to exchanges, and a superscript `i' a parameter relative
  to internal transformations.\relax
\mciteBstWouldAddEndPunctfalse
\mciteSetBstMidEndSepPunct{\mcitedefaultmidpunct}
{}{\mcitedefaultseppunct}\relax
\EndOfBibitem
\bibitem{balance}
It must be noted that the real activation and deactivation reactions are
  actually more complex than the simple reactions described in the model. The
  real reactions can easily be reduced to the simplified one. Typically, the
  implicit chemical compounds are such as water molecules, compounds resulting
  from acido-basic exchanges, or additional waste molecules having neglectable
  influence on the kinetics.\relax
\mciteBstWouldAddEndPunctfalse
\mciteSetBstMidEndSepPunct{\mcitedefaultmidpunct}
{}{\mcitedefaultseppunct}\relax
\EndOfBibitem
\end{mcitethebibliography}
}

\end{document}